\documentclass{article}

\usepackage{graphicx}
\usepackage{dcolumn}
\usepackage{bm}
\usepackage{color}
\usepackage{lscape}
\usepackage{amscd,amsmath,amssymb}
\newcommand{\be}{\begin{equation}}
\newcommand{\bel}{\begin{equation}\label}
\newcommand{\ee}{\end{equation}}

\newlength{\intwidth}

\begin{document}

\title{Remarks on the principles of statistical fluid mechanics}

\author{
  Koji Ohkitani\\
  Research Institute for Mathematical Sciences,\\
  Kyoto University,  Kyoto 606-8502 Japan.
}




\maketitle
\begin{abstract}
  This is  an idiosyncratic survey of statistical fluid mechanics
  centering on  the Hopf functional differential equation.
  Using the Burgers equation for illustration we review several functional integration approaches
  to theory of turbulence.
  We notice in particular  that some important contributions
have been brought about by researchers working on wave propagation in random media, among which
Uriel Frisch is not an  exception.
We also discuss a particular finite-dimensional approximation for the Burgers equation.

\end{abstract}



\begin{flushleft}
turbulence, Hopf equation, functional integration
\end{flushleft}

\section{Introduction}
In field theory one important task is the following.
Given an action integral $S[\phi]$, e.g. that of the $\phi^4$-model \cite{Edwards1964b}
$$S[\phi]=\iint dxdt \left\{ \frac{1}{2}\left( \phi_t^2 -|\nabla \phi|^2 \right)
-\frac{m^2}{2}\phi^2-\lambda\phi^4+J\phi\right\}$$
written with standard notations, we consider the partition function
$$Z(J)={\cal N} \int \exp \left( i S[\phi] \right) {\cal D}[\phi].$$
The objectives are to figure out how to make sense of what the symbol means mathematically
and to extract physically meaningful information from it.
In view of a correspondence between field theory and statistical mechanics, such a view
can be carried over to fluid mechanics. Hence in principle statistical theory of turbulence
should be based on that footing.

We will revisit the formulation of the Hopf functional differential equation (hereafter FDE)\cite{Hopf1952, HT1953}
which governs statistics of fluid turbulence. It is {\it not} our intention to give
an exhaustive survey of the subject matter. Rather we will
restrict our attention to one of its facets, that is, functional integration approaches
to theory of turbulence. In section 2 we compare some FDEs for statistical fluid mechanics with historical notes.
We  discuss decaying turbulence in Section 3 and forced turbulence in Section 4. In Section 5

\noindent 
we consider the Hopf equation for a discrete model. Section 6 is a brief summary.
This is  basically a survey article, except for Sections 4(b) and 5.


\section{The birth of the Hopf equation}

The primary objective of so-called ergodic theory is to show
that the ensemble average equals the time average for dynamical systems
of physical interest.
In 1930s there was substantial progress in the field,
including the mean ergodic theorem (strong in $L^2$) \cite{Neumann1929}
and  the individual ergodic theorem  (strong in $L^1$) \cite{Birkhoff1931}.
Hopf  extended von Neumann's work to weak convergence and published a monograph  \cite{Hopf1937}.
For its brief history, see e.g. \cite{TKS2013}.

In the development of $L^2$-theory an innocent-looking but far-reaching idea was put forward in \cite{Koopman1931}, that is,
the introduction of a kind of composition operators \cite{SM2012}. That enabled:
i) linearisation by considering observables and ii) spectral analysis due to unitarity of operators.
The price we have to pay is that we need to handle an infinite-dimensional space.
For an observable $f(\cdot)$ and a phase point $P$ the Koopman operator\footnote{The same concept
  was introduced independently by A.Weil \textcolor{black}{in connection with \cite{Weil1932}.
  See also \cite{Redei2005}, p.32.}
  Paying attention to an invariant measure in Hamiltonian systems,
  Weil realised that such a system defines a one-parameter group of unitary transformations in an $L^2$-space with respect
  to the measure.}
$U_t$ is defined by
\bel{Koopman}
U_t f(P)=f(\phi_t P),
\ee
where $\phi_t$ denotes a phase flow.
In the language of probability theory the Koopman operator is the adjoint group to the Perron-Frobenius operator 
and the process of  lifting of a dynamical system to a group of operators is called Koopmanism \cite{Applebaum2019}.
Hopf did not mention explicitly the notion of the Koopman operator in \cite{Hopf1952}, but  his
formulation of statistical fluid mechanics clearly rests on it, as the Hopf-Foias approach can be
regarded as such a manifestation.

Statistical study of the Navier-Stokes equations was initiated
on the basis of  a functional differential equation that now bears his name  \cite{Hopf1952, HT1953}. 
There the functional equation was heuristically derived for the characteristic functional of the velocity field.
On the other hand mathematical study of the Navier-Stokes equations was initiated in \cite{Foias1972, Foias1973}
on the basis of the  Liouville equation, a functional equation for the probability measure.
Further developments in this direction were summarised in \cite{VF1980}, including the so-called first integral method.
As a rule of thumb, Table \ref{Koopmanism} shows a rough classification of several representations in mechanics.
A related but different topic which we will not discuss here is the so-called Koopman mode analysis.
There has been significant progress recently therein and interested readers ought to consult
e.g.\cite{Mezic2013, RD2017, PK2018}.
\begin{table}
\begin{tabular}{|c|c|c|}
\hline
                    & quantum mechanics & fluid mechanics\\ \hline
group of operators (FDE) &  & $\stackrel{\mbox{Hopf-Foias formalism}}{\mbox{(or, Vishik-Fursikov formalism)}}$    \\ \hline
group of operators (PDE) & $\stackrel{\mbox{Heisenberg representation}^*}{\mbox{(or, Dirac representation)}}$ &   \\\hline
dynamical system & Schr{\"o}dinger equation  & Navier-Stokes equations \\
\hline
\end{tabular}
\caption{Some examples of Koopmanism in mechanics. $*$A rough analogy where evolution is described
  by time-dependent observables. }
\label{Koopmanism}
\end{table}

For the sake of simplicity, to illustrate the main ideas in this article  we will make use of the Burgers
equation in $\mathbb{R}^1$ 
\bel{Burgers}
\frac{\partial u}{\partial t} + u \frac{\partial u}{\partial x}=\nu \frac{\partial^2 u}{\partial x^2},
\ee
$$
u(x,0)=u_0(x),
$$
where $u_0(x)$  is a smooth initial condition.
They are transposable at least formally to the Navier-Stokes equations. Thus, when we say turbulence in this paper
we mean a flow of the 1D model.

Hopf considered the characteristic functional of the velocity field
$\Phi[\theta(x),t]=\left< \exp i \int u(x,t)\theta(x) dx\right>,$ where
$\left<\cdot\right>$ denotes an ensemble average with respect to $u_0$.
The governing equation he derived in \cite{Hopf1952} reads 
\bel{Hopf}
\frac{\partial \Phi}{\partial t}=
\frac{i}{2}\int  \theta(x) \frac{\partial}{\partial x}
\frac{\delta^2 \Phi}{\delta \theta(x)dx^2} dx
+\nu \int \theta(x) \frac{\partial^2}{\partial x^2}
\frac{\delta \Phi}{\delta \theta(x) dx} dx.
\ee
This is a functional differential equation (FDE). \textcolor{black}{
We recall that when the functional varies as $\delta \Phi[\theta(x)] \approx \int A(x)\delta \theta(x) dx$
for a small variation $\delta \theta(x),$     
the functional derivative $\frac{\delta \Phi}{\delta \theta(x) dx}$ is defined  by $A(x)$.}

Let us write down back to back the Liouville equation, which is also known as the Hopf-Foias equation 
\bel{Liouville}
\frac{d}{dt}\int \Psi[u] d\mu_t(u)
=\int \left( \nu \frac{\partial^2 u}{\partial x^2}- u \frac{\partial u}{\partial x}, \, \frac{\delta \Psi}{\delta u(x)dx} \right)  d\mu_t(u).
\ee
This can be derived as follows\cite{Foias1974}.
Define a mapping $T: u \longrightarrow u+ (\nu u_{xx} -u u_x )dt$ and consider an observable $\Psi[u]$.
By  $\mu_{t + dt}=\mu_t \circ T^{-1}$ (Hopf identity), we have $\int \Psi[u] d\mu_{t+dt} =\int (\Psi \circ T)d \mu_t.$
Now we have
\be\mbox{LHS} \approx  \int \Psi[u] d\mu_{t}+
\frac{d}{dt}\left(\int \Psi[u] d\mu_{t} \right) dt,\nonumber
\ee
whereas
\begin{eqnarray}
\mbox{RHS} &\approx& 
\int \Psi\left[ u +\left(\nu \frac{\partial^2 u}{\partial x^2}- u \frac{\partial u}{\partial x}\right)dt\right]  d\mu_{t}\nonumber\\
&\approx& \int \Psi[u] d\mu_{t}+
\int \left(\nu \frac{\partial^2 u}{\partial x^2}
- u \frac{\partial u}{\partial x},\, \frac{\delta \Psi}{\delta u(x)dx} \right) d\mu_{t} \,dt. \nonumber
\end{eqnarray}
Hence the Liouville equation follows. $\square$\\
It should be noted that this equation is valid for any observables. In particular,
to obtain the original form of the Hopf functional equation we may
simply take $\Psi[u]=\exp \left(i (u,\theta) \right),$ where $(u,\theta)$ denotes an inner-product,
$(u,\theta)=u \cdot \theta=\int u(x,t)\theta(x)dx.$

We note in passing that the equation for probability density function $F=\Pi_k \delta(u_k -u_k(x,t))$
was written down in \cite{Edwards1964a, Edwards1964b}, where the symbolic product over $k$ denotes the one
over all $x$.
Adapting to the Burgers equation, it takes the following form
\bel{Edwards}
\frac{\partial F}{\partial t}=\int \frac{\delta}{\delta u(x)\textcolor{black}{dx}}
\left(\nu \frac{\partial^2 u}{\partial x^2} - u \frac{\partial u}{\partial x}\right)\,F dx.
\ee
This is equivalent to the Hopf-Foias equation (\ref{Liouville}).
Anyway in the Liouville-type equations the underlying PDE appears explicitly. A complaint
made by mathematical analysts about the Hopf equation is that they don't see PDEs in it.
Actually PDEs do appear in the path integral representation of their formal solution of the latter.

\section{Decaying turbulence}

\subsection{Rosen's action integral}

Apparently the first path integral representation of this ilk is due to \cite{Rosen1960}.
We begin considering freely decaying turbulence.
A formal and symbolic  solution was constructed to the Hopf equation in \cite{Rosen1960}.
See also \cite{Rosen1969}. 
Making use of the linearity of the Hopf equation we write the solution as
\bel{semigroup}
\Phi[\theta(x),t]=\int \Phi_0 [\theta_0(x)]  K[\theta_0(x)|\theta(x),t]{\cal D}[\theta_0(x)],
\ee
for a Green's function $K[\theta_0(x)|\theta(x),t].$ We will make use of the \textcolor{black}{plane} wave expansion of the Dirac mass
\bel{delta}
\delta(x)=\frac{1}{2\pi}\int_{-\infty}^{\infty} \exp (ikx) dk,
\ee
or more precisely, in its functionally generalised form \cite{Novikov1961}
\bel{Delta}
\delta[\psi]=\frac{1}{2\pi}\int \exp \left\{ i(\psi\cdot\phi) \right\} {\cal D}[\phi].
\ee
A symbolic expression of the Green's function $K[\theta_0(x)|\theta(x),t]$
was given as follows:
\bel{propagator}
K[\theta_0(x)|\theta(x),t]=
C\iint_{\eta(x,0)=\theta_0(x)}^{\eta(x,t)=\theta(x)}
\exp \left\{ i \int_0^t d\tau \int
\left(\frac{\partial \eta}{\partial \tau}\zeta + \eta Q[\zeta]\right) dx
\right\}{\cal D}[\eta(x,\tau)]{\cal D}[\zeta(x,\tau)],
\ee
where $Q[\zeta]=-\zeta \zeta_x+\nu \zeta_{xx}$ and $C$ is a formal normalization constant.
In those expressions e.g. ${\cal D}[\zeta]=\Pi_{x} d\zeta(x,t)$ denotes a fictitious (translation-invariant)
measure.\footnote{We may call it 'abstract Feynman measure'. It may be compared with the  classical Wiener measure
  in $\mathbb{R}^n$  and the abstract Wiener measure in $\mathbb{R}^{\infty},$ both of which are well understood.
  Abstract Feynman measure does not exit because Feynman measure does not, see Appendix.}

Actually the basic formula (\ref{Delta}) is common to all the derivations we review in this article.

We recall how Rosen's  path integral representation is derived.
Because of linearity it suffices to show that $K$ satisfies the Hopf equation.
For small $\Delta t$ the Green's function is proportional to
\begin{eqnarray}
& &  \int \exp \left\{ i \int \left( (\eta-\eta')\cdot \zeta +\Delta t \eta Q[\zeta] \right)dx\right\} {\cal D}[\zeta]\nonumber\\
  &=&\int \exp \left\{ i \int \zeta \cdot ( \eta-\eta') dx \right\}\exp \left\{  i \Delta t \int \eta(x) Q[\zeta] dx \right\} {\cal D}[\zeta]\nonumber\\
&\approx& \int \left( 1+ i \Delta t \int \eta(x) Q[\zeta] dx \right)
\exp \left\{ i \int \zeta \cdot(\eta-\eta') dx \right\} {\cal D}[\zeta]\nonumber\\
&=& \left( 1+ i \Delta t \int dx \eta(x) Q\left[ \frac{\delta}{i\delta \eta}\right]  \right)
\int  \exp \left\{ i \int \zeta \cdot ( \eta-\eta') dx  \right\} {\cal D}[\zeta],\nonumber
\end{eqnarray}
where we have written $\eta=\eta(x,t+\Delta t),\eta'=\eta(x,t).$
Hence we have by (\ref{delta})
$$K(t|t+\Delta t)= \left( 1+ i \Delta t \int dx \eta(x) Q\left[ \frac{\delta}{i\delta \eta}\right]  \right)
\underbrace{\delta[\eta-\eta']}_{=K(t|t)},$$
that is,
$$\frac{\partial K}{\partial t}=i\int  \eta(x) Q\left[ \frac{\delta}{i\delta \eta}\right] Kdx.\;\square$$

Out of the double functional integrations, the ${\cal D}[\zeta]$ integration can be carried out
by the method of stationary phase \cite{Rosen1960}, but the ${\cal D}[\eta]$ integration remains a challenge\footnote{If it were possible to carry out the ${\cal D}[\eta]$ integration, the resultant functional would be termed the Onsager-Machlup action in analogy with similar results for statistics of linear dynamical systems. e.g. Section VI.9 of \cite{IW2014}.} \cite{Tatarskii1962, MY1975}.

For the particular case of the heat equation, at least we know
$$C\iint_{\eta(x,0)=\theta_0(x)}^{\eta(x,t)=\theta(x)}
\exp \left\{ i \int_0^t d\tau \int 
\zeta\left(\frac{\partial \eta}{\partial \tau} + \nu \frac{\partial^2 \eta}{\partial x^2}\right)dx
\right\}{\cal D}[\eta(x,\tau)]{\cal D}[\zeta(x,\tau)]
=\delta[\theta_0 -\exp(\nu t \triangle) \theta]$$
so that we recover
$\Phi[\theta, t]=\Phi_0[\exp(\nu t \triangle)\theta].$
This is a solution which represents the final period of decay in turbulence \cite{Hopf1952, HT1953}.

\subsection{Alternative derivation for Rosen's path integral}
\textcolor{black}{Rosen's original derivation of the characteristic functional based on the Hopf equation is
mostly straightforward.}
We recall that Rosen's symbolic expression  can be obtained \textit{without}
using the Hopf equation, which is allegedly based on the idea of Novikov, Section 29.5 of \cite{MY1975}.
See also \cite{Klyatskin2015}, p.252.
\textcolor{black}{The rationale for including this is that the second derivation stands valid when
  we have external forcing which is not necessarily white-in-time Gaussian.
  For simplicity we illustrate it here using the unforced equation.}  

We formally rewrite $\frac{\partial u}{\partial t}=Q[u]$  as
$$u(x,t)=u_0(x)+\int_{0}^{t} Q[u(x,\tau)]d\tau.$$
In fact, this virtually trivial recasting leads to a nontrivial result.
Consider
$$\exp\left( i \theta(x) \cdot u(x,t) \right)
= \exp\left\{ i \theta(x) \cdot \left( u_0(x)+\int_{0}^{t} Q[u(x,\tau)] d\tau \right)\right\}$$
$$=\iint \exp\left\{ i \theta(x)\cdot \left( u_0(x)+\int_{0}^{t} Q[v(x,\tau)]d\tau \right)\right\}
\exp\left( i\int_{0}^t  (u(x,\tau)-v(x,\tau)) w(x,\tau) d\tau  \right) {\cal D}[v]{\cal D}[w],
$$
by virtue of (\ref{Delta}).
Plugging 
$u(x,\tau)=u_0(x)+\int_{0}^{\tau} Q[v(x,\tau')]d\tau'$
into the above equation, we have
$$\exp\left( i \theta(x) \cdot u(x,t) \right)
=\iint \exp\left\{ i \theta(x) \cdot \left( u_0(x)+\int_{0}^{t} Q[v(x,\tau)]d\tau \right)\right\}$$
$$\times \exp\left\{ i\int_{0}^t  \left(u_0(x)+\int_{0}^{\tau} Q[v(x,\tau')]d\tau' -v(x,\tau)\right) w(x,\tau)
d\tau  \right\} {\cal D}[v]{\cal D}[w].$$
Integrating by parts we find
$$\int_{0}^t d\tau \int_{0}^{\tau} Q[v(x,\tau')]d\tau'  w(x,\tau)
=\left[\int_t^{\tau} w(x,\tau') d\tau' \int_{0}^\tau Q[v(x,\tau')]d\tau' \right]_{\tau=0}^t
-\int_{0}^t d\tau \int_t^{\tau}  w(x,\tau') d\tau' Q[v(x,\tau)]$$
$$
=+\int_{0}^t d\tau \int_{\tau}^t w(x,\tau') d\tau' Q[v(x,\tau)],
$$
and hence deduce
$$e^{i(u,\theta)}=\iint \exp \left\{i u_0(x) \cdot\left( \theta(x) +  \int_{0}^t w(x,\tau') d\tau'\right)\right\}
\exp \left\{i \int_{0}^t d\tau Q[v]\cdot\left( \theta(x) +  \int_{\tau}^t w(x,\tau') d\tau'\right)\right\}$$
$$\times \exp \left\{-i \int_{0}^t  v (x,\tau)\cdot w(x,\tau) d\tau \right\}  {\cal D}[v]{\cal D}[w].$$
Averaging over $u_0$, we find
$$\Phi[\theta(x),t]=\iint
\Phi_0  \left[\theta(x) +  \int_{0}^t w(x,\tau') d\tau'\right]$$
$$\times \exp  \left\{i \int_{0}^t d\tau \left( Q[v]\cdot\left( \theta(x) +  \int_{\tau}^t w(x,\tau')d\tau'\right)
- v (x,\tau)\cdot w(x,\tau) \right) \right\}  {\cal D}[v]{\cal D}[w].
$$
We rename $v(x,\tau)=\zeta(x,\tau)$ and $\theta(x)+ \int_{\tau}^t w(x,\tau')d\tau'= \eta(x,\tau)$ with
$\eta(x,t)=\theta(x), \eta(x,0)=\theta(x) +\int_{0}^t w d\tau'$ and $d\eta(x,\tau)=-w(x,\tau)d\tau.$
Replacing ${\cal D}[v] = {\cal D}[\eta]$ and ${\cal D}[w] \propto {\cal D}[\eta]$ and subsuming
extra factors into the prefactor,
we recover (\ref{semigroup}, \ref{propagator}) with the lower limit of integration unspecified. $\square$

\section{Stationary turbulence}
\subsection{Generalised Hopf equation}
We now turn our attention to statistically steady turbulent flows driven by an external forcing.
The Burgers equation with the forcing reads
$$\frac{\partial u}{\partial t} +u\frac{\partial u}{\partial x}=\nu\frac{\partial^2 u}{\partial x^2}
+f(x,t).$$
For simplicity let us consider a white-in-time Gaussian random force $f(x,t)$ \cite{Edwards1964a},
which satisfies
$$\left<f(x,t)f(y,s) \right>=F(x-y)\delta(t-s).$$
Here $F(r)$ denotes the correlation function and $\left< \cdot \right>$  a statistical average. 
That way we can decouple the joint probability distributions of the velocity and the forcing.

The Hopf equation was generalized to accommodate random forcing in \cite{Novikov1965}.
This is based on the following Furutsu-Novikov-Donsker theorem \cite{Novikov1965, Furutsu1963,
Donsker1964, DL1962}, which states for any functional $R$
\bel{FND}
\left< f(x)R[f] \right>
=\int F(x-x') \left< \frac{\delta R[f]}{\delta f(x') dx'} \right>dx'.
\ee
The generalised Hopf equation turned out to be
\bel{Novikov}
\frac{\partial \Phi}{\partial t}=
\frac{i}{2}\int  \theta(x) \frac{\partial}{\partial x}
\frac{\delta^2 \Phi}{\delta \theta(x) dx^2} dx
+\nu \int \theta(x) \frac{\partial^2}{\partial x^2}
\frac{\delta \Phi}{\delta \theta(x)dx} dx
-\frac{1}{2}\iint \theta(x) F(x-y) \theta(y)dx dy\,\Phi.
\ee
For the Hopf equation subject to more general forcing, see \cite{Hosokawa1968}. 

Just like the original Hopf equation, it seems that this equation has not been put to practical use.
Here let us see how we may modify Rosen's functional expression to accommodate the forcing
and verify that it satisfies the generalized Hopf equation.
To that end the following paragraph by Uriel Frisch would be instructive,
which we quote here \textit{verbatim} from \cite{Frisch1968}.\\

{\em
We consider first the randomly perturbed heat equation
\be
\frac{\partial}{\partial t}\Psi(t,\bm{r})=\triangle \Psi+\mu(\bm{r})\Psi,\;\;\Psi(0,\bm{r})=\delta(\bm{r}),\tag{3.44}
\ee
where $\mu(\bm{r})$ is a centered random Gaussian function with covariance
\be
\Gamma(\bm{r},\bm{r}')={\cal E}\left\{\mu(r)\mu(r')\right\}. \tag{3.45}
\ee
For any realization of $\mu(\bm{r}),$ Eq.(3.44) can be solved by means of Kac's formula [cf. Appendix Eq. (A12)]
\be
\Psi(t,\bm{r})={\cal E}_W\left\{ \delta(\bm{\rho}(t)-\bm{r})\exp \left\{\int_{0}^{t}\mu(\bm{\rho}(\tau))d\tau \right\}\right\}.\tag{3.46}
\ee
To find ${\cal E}_\mu(\Psi(t,\bm{r}))$ we must average again over $\mu$. Assuming that the $\mu$ and $W$ averages can be interchanged,
we obtain
\be
{\cal E}_\mu\{\Psi(t,\bm{r})\}={\cal E}_W\left\{ \delta(\bm{\rho}(t)-\bm{r}) {\cal E}_{\mu}
\left\{\exp \left\{\int_{0}^{t}\mu(\bm{\rho}(\tau))d\tau \right\}\right\}\right\}. \tag{3.47}
\ee
For a fixed Brownian path $\bm{\rho}(\tau),\;0\leq \tau \leq t$, the random variable
\be
\phi=\int_0^t\mu(\bm{\rho}(\tau))d\tau \tag{3.48}
\ee
is a linear functional of the Gaussian random function $\mu$; hence
\be
{\cal E}_\mu\{ \exp\phi \}= \exp \left\{\frac{1}{2} {\cal E}_\mu\{\phi^{2}\} \right\}
=\exp\left\{ \frac{1}{2}\int_0^t \int_0^t \Gamma(\bm{\rho}(\tau),\bm{\rho}(\tau'))d\tau d\tau' \right\}. \tag{3.49}
\ee
Eq.(3.47) can now be rewritten as
\be
{\cal E}_\mu\{\Psi(t,\bm{r})\}={\cal E}_W\left\{ \delta(\bm{\rho}(t)-\bm{r})
\exp\left\{ \frac{1}{2}\int_0^t \int_0^t \Gamma(\bm{\rho}(\tau),\bm{\rho}(\tau'))d\tau d\tau' \right\}\right\}. \tag{3.50}
\ee
}
\subsection{Linking Rosen's action integral to Wyld's}
Note that
the action integral for turbulence driven by Gaussian white forcing is
called Wyld functional\cite{Zinn-Justin2021} and appears in the so-called Martin-Siggia-Rose formalism (hereafter, MSR)
\cite{MSR1973}.
The formalism builds a field theory associated with statistical mechanics of a given dynamical system.
That appears in a now standard field-theoretic treatment of turbulence, particularly
instanton theory. Such an action integral is sometimes called after the names of Wyld\footnote{The explicit functional form,
however, does not appear explicitly in \cite{Wyld1961}.} and Rosen  e.g. \cite{BdS1998}.
The link is well expected \textcolor{black}{in the folklore}, but best described here explicitly with some justifications.

\textcolor{black}{
Before that it is suitable to refer some cornerstone papers on the theory of turbulence. The space-time characteristic
functional was introduced in \cite{LK1962}, the path integral representation for the MSR theory was derived
in \cite{Dominicis1976, Janssen1976} and a connection to the Kolmogorov theory was sought in \cite{DM1979}.
See also Section 9.5 of \cite{Frisch1995} for a succinct survey.}

We will see  that (i) the symbolic expression associated with Rosen's action integral subject to random forcing
yields with the MSR action and (ii) it is indeed a solution of Novikov's generalised Hopf equation.
To include forcing we replace RHS of the Burgers equation
$Q[\zeta] \rightarrow Q[\zeta] +f(x,t)$ where $Q[\zeta]=-\zeta \zeta_x+\nu \zeta_{xx},$
for each realization of $f(x,t)$, that is, we regard $f(x,t)$ as deterministic at this stage.
We have
$$K[\theta_0(x)|\theta(x),t]$$
$$
=C\iint_{\eta(x,0)=\theta_0(x)}^{\eta(x,t)=\theta(x)}
\exp \left\{ i \int_0^t d\tau \int dx\,
\left(\frac{\partial \eta}{\partial \tau}\zeta + \eta Q[\zeta]\right)\right\}
\exp \left\{ i \int_0^t d\tau \int \eta(x)f(x) dx\,\right\}
    {\cal D}[\eta(x,\tau)]{\cal D}[\zeta(x,\tau)].$$
    Randomise $f(x,t)$ and average over ${\cal D}[f(x,\tau)]$ with the weight
    $\exp \left\{-\frac{1}{2}\int_0^t d\tau\iint f(x) \kappa (x-y) f(y)dx dy \right\}$
    where $\kappa(x,y)=\kappa(x-y)$ is an inverse
    of $F(x,y)=F(x-y).$
    We find
$$K=C\iint_{\eta(x,0)=\theta_0(x)}^{\eta(x,t)=\theta(x)}
\exp \left\{ i \int_0^t d\tau \int dx\,
\left(\frac{\partial \eta}{\partial \tau}\zeta + \eta Q[\zeta]\right)
 -\frac{1}{2} \int_0^t d\tau \iint \eta(x) F(x-y) \eta(y)dx dy \right\}
    {\cal D}[\eta]{\cal D}[\zeta].$$
  
    It can be seen to be equivalent to the Wyld functional in the MSR scheme. First,
    integrating by parts under the constraints $\zeta(x,0)=\zeta(x,t)=0$,  we have
    $$K=C\iint_{\eta(x,0)=\theta_0(x)}^{\eta(x,t)=\theta(x)}
\exp \left\{- i \int_0^t d\tau \int dx\,
\eta \left(\frac{\partial \zeta}{\partial \tau} - Q[\zeta]\right)
 -\frac{1}{2} \int_0^t d\tau \iint \eta(x) F(x-y) \eta(y)dx dy \right\}
    {\cal D}[\eta]{\cal D}[\zeta].$$
Second, to verify that the expression defined with the above kernel satisfies the generalised Hopf equation we note
$$\frac{\partial}{\partial t}\exp \left\{ -\frac{1}{2} \int_0^t d\tau \iint \eta(x,\tau) F(x-y) \eta(y,\tau)dx dy \right\}$$
$$  =  -  \frac{1}{2} \iint \eta(x,t) F(x-y) \eta(y,t)dx dy 
    \exp \left\{ -\frac{1}{2} \int_0^t d\tau \iint \eta(x) F(x-y) \eta(y)dx dy  \right\}.$$
The additional contribution to the RHS of the Hopf equation is
$$-\frac{C}{2}\iint_{\eta(x,0)=\theta_0(x)}^{\eta(x,t)=\theta(x)}
\iint \left. \eta(x,t) F(x-y) \eta(y,t)dx dy \right|_{\eta(\cdot,t)=\theta(\cdot)}$$
$$ \times \exp \left\{ -i \int_0^t d\tau \int dx\,
\eta \left(\frac{\partial \zeta}{\partial \tau} - Q[\zeta]\right)
 -\frac{1}{2} \int_0^t d\tau \iint \eta(x) F(x-y) \eta(y)dx dy \right\}
    {\cal D}[\eta]{\cal D}[\zeta].$$
    $$=-\frac{1}{2} \iint \theta(x) F(x-y) \theta(y)dx dy$$
$$\times  C   \iint_{\eta(x,0)=\theta_0(x)}^{\eta(x,t)=\theta(x)}\exp \left\{- i \int_0^t d\tau \int dx\,
    \eta \left(\frac{\partial \zeta}{\partial \tau} - Q[\zeta]\right)
  -\frac{1}{2} \int_0^t d\tau \iint \eta(x) F(x-y) \eta(y)dx dy \right\}
    {\cal D}[\eta]{\cal D}[\zeta],$$
which indeed yields the final term on the RHS of Novikov's equation (\ref{Novikov}). $\square$

\subsection{Instanton theory}
The functional integrations remain merely symbolic. However, the action integral that
appears in the exponent is just $(n+1)$-dimensional integral and well-defined $(n=1,2,3)$.
Hence it is sensible to consider stationary conditions by writing down the Euler-Lagrange equations.
For Rosen's action, a pair of equations consists of the Burgers equation and its adjoint (dual)
$$
\frac{\partial v}{\partial t} + u \frac{\partial v}{\partial x}=-\nu \frac{\partial^2 v}{\partial x^2},
$$
which is linear and runs backward in time. Statistical fluid mechanics
has inherently control-theoretic aspects built-in \cite{GGS2015}.

Here we will have a look at standard manipulations in instanton theory, following \cite{GM1996,FKLM1996}.
See also \cite{GGS2015, GGSV2015}for the numerical implementations, including the Navier-Stokes equations.

Consider a formal normalisation
$ {\cal N}=\int \delta(u_t+uu_x-\nu u_{xx}-f)  {\cal D}[u] $
and the partition function
$$Z=\int \exp \left( -\frac{1}{2} \iiint f(x,t)\kappa(x-y)f(y,t) dx dy dt\right) {\cal D}[f]$$
$$=\frac{1}{\cal N}\iint \delta(u_t+uu_x-\nu u_{xx}-f) \exp \left( -\frac{1}{2} \iiint f(x,t)\kappa(x-y)f(y,t) dx dy dt\right)
{\cal D}[f] {\cal D}[u].$$
We rewrite this by (\ref{Delta}), redefining the normalisation constant
$$Z=\frac{1}{\cal N}\iiint  \exp \left( i\iint v (u_t+uu_x-\nu u_{xx}-f) dxdt
-\frac{1}{2}\iiint f(x,t)\kappa(x-y)f(y,t) dx dy dt\right) {\cal D}[f] {\cal D}[u] {\cal D}[v]$$
$=\frac{1}{\cal N}\iiint  \exp \left( i\iint v (u_t+uu_x-\nu u_{xx}) dxdt \right)$
$$\times \exp \left( -i \iint  v f dxdt -\frac{1}{2}\iiint f(x,t)\kappa(x-y)f(y,t) dx dy dt\right)
{\cal D}[f] {\cal D}[u] {\cal D}[v]$$
$$=\frac{1}{\cal N}\iint  \exp \left( i\iint v (u_t+uu_x-\nu u_{xx}) dxdt \right)
\exp \left(  -\frac{1}{2}\iiint v(x,t)F(x-y)v(y,t) dx dy dt\right)  {\cal D}[u] {\cal D}[v],$$
where a functional Fourier transform of the Gaussian function has been taken in the final line.
\subsection{Yet another form of path integral}
A simple conversion of the Burgers equation into an integral equation was
considered in 3(b). By virtue of Duhamel principle we can also convert
the Burgers equation into another integral equation of the following form
$$u-G:uu={\hat u},$$
where
${\hat u}=g*u_0$ is a heat flow defined with the heat kernel
$g(x,t)=\dfrac{1}{(4\pi \nu t)^{1/2}}\exp \left(-\dfrac{x^2}{4 \nu t}\right).$\\
\noindent We have also denoted
$$G:uu \equiv \int_0^t ds \int_{-\infty}^{\infty} dy G(x-y,t-s)u(y,s)^2,$$
where $G(x,t) \equiv -\frac{1}{2}\partial_x g(x,t).$
On this basis the following representation can be derived as in \cite{Rosen1971, TR1974}
\bel{Rosen_no2a}
\Phi[\theta]= \iint \exp \left(i \int \left\{(\theta-v)\cdot u
+v\cdot G:uu\right\} dx -A[u]
\right){\hat\Phi}[v] {\cal D}[v] J {\cal D}[u],
\ee
where $A[u]$ denotes an unknown functional to account for possible breakdown of classical solutions
and $J={\rm det} \left( \frac{\delta{\hat u}}{\delta u}\right).$

In fact we can confirm that $J\equiv 1$ as in the 3D incompressible case. We also have $A[u]\equiv 0$
because the Burgers equation is globally well-posed. Hence we find
\bel{Rosen_no2b}
\Phi[\theta]= \iint \exp \left(i \int \left\{(\theta-v)\cdot u
+v\cdot G:uu\right\} dx
\right){\hat\Phi}[v] {\cal D}[v]  {\cal D}[u].
\ee
This formula connects the characteristic functional ${\hat\Phi}[v]$ of the heat equation
with that of the Burgers equation $\Phi[\theta]$.
If we discard the second term of the exponent, it is clear that we recover
$\Phi[\theta]={\hat\Phi}[v]$ by the property of the Dirac measure (\ref{Delta}).
This form of path integral is valid only for a viscous fluid, in contrast to the previous one
which can cover an inviscid fluid as well.
It is of interest to study whether and how the above expression is simplified further
when we make use of the Cole-Hopf transform which linearises the Burgers equation.

\section{Hopf equation for a discrete Burgers  equation}
As seen above the Hopf functional equation faces a daunting mathematical difficulty
because of the infinite-dimensional character of functional differential equation.
To alleviate the difficulty at least partially,
we consider a  discrete Burgers  equation in this section so that its 'Hopf equation'
is a PDE rather than an FDE. 

There are some works in the similar spirit \cite{BP1993, BP1996a, BPW1996b}.
The master equation  was presented for a discrete Burgers equation, but that
does not respect Cole-Hopf transform. See also \cite{Kollmann1976}.
Here we take up a discretisation of Burgers equation which respects the Cole-Hopf transformation
and present  asymptotic analysis using a series expansion in time.

\subsection{Discrete Burgers equation}
The following version of discretisation was proposed in \cite{Hirota1979} for the Burgers equation.
\bel{Hirota1}
\frac{d u(x,t)}{d t}
=-\frac{2\nu^2}{\epsilon^2}\Delta_x
\left\{ \exp \left( -\frac{\epsilon}{2\nu} u(x,t)  \right)
   +\exp \left( \frac{\epsilon}{2\nu} u(x-\epsilon,t)  \right) -2
   \right\}
\ee
\be
   =-\frac{2\nu^2}{\epsilon^3}
\left\{ \exp \left( -\frac{\epsilon}{2\nu} u(x+\epsilon,t)  \right)
-\exp \left( -\frac{\epsilon}{2\nu} u(x,t)  \right)
+\exp \left( \frac{\epsilon}{2\nu} u(x,t)  \right)
   -\exp \left( \frac{\epsilon}{2\nu} u(x-\epsilon,t)  \right) 
  \right\}, \nonumber
\ee
  where $\epsilon$ denotes spatial spacing and $\Delta_x f(x)=(f(x+\epsilon)-f(x))/\epsilon.$
Writing $u_n(t)=u(x_n,t)$ with $x_n=n\epsilon,\;(N=1,\ldots,N),$ it can also be written  
\be
\frac{d u_n}{dt}
=-\frac{2\nu^2}{\epsilon^3}\Delta_n
\left\{ \exp \left(-\frac{\epsilon}{2\nu} u_{n} \right)
+\exp \left(\frac{\epsilon}{2\nu}u_{n-1}\right)
-2 \right\},
\ee
where  $\Delta_n f_n =f_{n+1}-f_n$ and periodic boundary conditions  $f_N=f_0, f_{N+1}=f_1$
are assumed.
 \subsection{Its Hopf equation}
Consider the characteristic function 
$\Phi[\theta_n]=\left< \exp i\sum_{n=1}^N u_n(t) \theta_n \right>,$
then its governing equation can be derived as
\bel{Hopf_Hirota}
\frac{\partial \Phi}{\partial t}=-\frac{2\nu^2 i}{\epsilon^3}
\sum_{n=1}^N \theta_n \Delta_n \left( 
\Phi \left[\theta_1, \ldots, \theta_{n}+\frac{\epsilon i}{2\nu}, \ldots,\theta_N \right] 
+\Phi \left[\theta_1, \ldots, \theta_{n-1}-\frac{\epsilon i}{2\nu}, \ldots,\theta_N \right] 
-2\Phi \left[\theta_1, \ldots, \theta_N \right] 
\right).
\ee
It is readily checked that (\ref{Hopf_Hirota}) reduces to (\ref{Hopf}) in the limit of small
$\epsilon$.
Using exponential operators we can recast it as 
$$\frac{\partial \Phi}{\partial t}
 =-\frac{2\nu^2 i}{\epsilon^3}
 \sum_{n=1}^N \theta_{n} \Delta_n 
 \left\{ \exp\left(
 \frac{\epsilon i}{2\nu}\partial_{n} \right)
+\exp\left(-\frac{\epsilon i}{2\nu}\partial_{n-1} \right) -2
\right\}\Phi,
$$
$$
 =\frac{2\nu^2 i}{\epsilon^2}
 \sum_{n=1}^N (\widetilde{\Delta_n}\theta_{n-1})
 \left\{ \exp\left(
 \frac{\epsilon i}{2\nu}\partial_{n} \right)
+\exp\left(-\frac{\epsilon i}{2\nu}\partial_{n-1} \right) -2
\right\}\Phi,
$$
where  the final line follows from 'summation by parts' and
$\widetilde{\Delta_n} f_n = \frac{1}{\epsilon}\Delta_n f_n,\,
\theta_0 = \theta_N, \theta_{N+1}=\theta_1$ (by periodicity).
We can then write its formal solution
\bel{Hopf_sol}
\Phi=\exp\left[
\frac{2\nu^2 i t}{\epsilon^2}
 \sum_{n=1}^N (\widetilde{\Delta_n}\theta_{n-1})
 \left\{ \exp\left(
 \frac{\epsilon i}{2\nu}\partial_{n} \right)
+\exp\left(-\frac{\epsilon i}{2\nu}\partial_{n-1} \right) -2
\right\}
\right]  
\Phi_0[\theta_1,\ldots,\theta_N].
\ee
Expanding in a power series in time, we have
$$\Phi=\sum_{k=0}^\infty \frac{1}{k!} \left( \frac{2\nu^2 i t}{\epsilon^2} \right)^k
\left[ \sum_{n=1}^N (\widetilde{\Delta_n}\theta_{n-1})
 \left\{ \exp\left(
 \frac{\epsilon i}{2\nu}\partial_{n} \right)
+\exp\left(-\frac{\epsilon i}{2\nu}\partial_{n-1} \right) -2
\right\}\right]^k \Phi_0[\theta_1,\ldots,\theta_N].$$
These operators are non-commutative, e.g.
$$\exp\left(\frac{\epsilon i}{2\nu}\partial_{n} \right)(\theta_n-\theta_{n-1})
=(\theta_n-\theta_{n-1}) \exp\left(\frac{\epsilon i}{2\nu}\partial_{n} \right)+\frac{\epsilon i}{2\nu},$$
but the non-commutativity appears in higher order in $\epsilon$.
Hence, by multinomial expansion, we find to leading order 
 $$[\ldots]^k \Phi_0 \approx
\sum_{\sum_{j=1}^N k_j=k,\,k_j \geq 0}\frac{k!}{k_1! k_2! \ldots k_N!}
\Pi_{j=1}^N \left( \widetilde{\Delta_j}\theta_{j-1} \right)^{k_j}
\left\{ \exp\left(\frac{\epsilon i}{2\nu}\partial_{j} \right)                                                                 
+\exp\left(-\frac{\epsilon i}{2\nu}\partial_{j-1} \right) -2 
\right\}^{k_j}\Phi_0,$$
Again, by multinomial expansion
$$\left\{\ldots\right\}^{k_j}\Phi_0
=\sum_{\sum_{j=1}^3 m_j=m,\; m_j \geq 0} \frac{k_j !}{m_1! m_2! m_3!}\exp\left(\frac{i\epsilon }{2\nu}m_1\partial_{j}\right)
\exp\left(-\frac{i\epsilon }{2\nu}m_2\partial_{j-1}\right) (-2)^{m_3} \Phi_0[\theta_1,\ldots,\theta_N]
$$
$$
=\sum_{\sum_{j=1}^3 m_j=m,\; m_j \geq 0} \frac{k_j ! (-2)^{m_3}}{m_1! m_2! m_3!}
 \Phi_0\left[\theta_1,\ldots,\theta_{j-1} -\frac{i\epsilon }{2\nu}m_2,\theta_j+\frac{i\epsilon }{2\nu}m_1,\ldots,\theta_N\right].
 $$
 All together we obtain
$$\Phi \approx \sum_{k=0}^\infty \frac{1}{k!} \left( \frac{2\nu^2 i t}{\epsilon^2} \right)^k
\sum_{\sum_{j=1}^N k_j=k,\,k_j \geq 0}\frac{k!}{k_1! k_2! \ldots k_N!}
\Pi_{j=1}^N \left( \widetilde{\Delta_j}\theta_{j-1} \right)^{k_j}$$
$$\times \sum_{\sum_{j=1}^3 m_j=m,\; m_j \geq 0} \frac{k_j ! (-2)^{m_3}}{m_1! m_2! m_3!}
 \Phi_0\left[\theta_1,\ldots,\theta_{j-1} -\frac{i\epsilon }{2\nu}m_2,\theta_j+\frac{i\epsilon }{2\nu}m_1,\ldots,\theta_N\right].
$$
Note that in the above expressions $j$-dependence of $m_1,m_2,m_3$ has been suppressed for simplicity. 

\section{Summary}
After making a note on the Koopman (and Weil) operator, in this survey we reviewed some functional
integration
methods for handling the Hopf equation for turbulence. We have seen a number of different ways of path
integral representations for both decaying turbulence (with Rosen's action) and forced turbulence (with
Wyld's action).
We have also seen how those two actions are related via a standard technique known to researchers
on wave propagation, as evidenced by Uriel's paragraph.
The technique is also useful in modern instanton theory.
Common to all those derivations lies the \textcolor{black}{plane} wave expansion of the Dirac delta measure,
which is essentially a tool in linear theory (Fourier analysis).

In the final part we discuss a particular finite-dimensional approximation for the Burgers equation
and construct approximate solutions to its Hopf characteristic function.

For recent references on statistical theory of turbulence from functional integration viewpoints,
it may be in order to consult
e.g. \cite{Botelho2008, Klyatskin2015, Kollmann2019, Venturi2018, Ohkitani2020}.
For the developments of turbulence theory in general, the following references are also of interest 
\cite{SE2005, EF2011}.\\

A very happy birthday, Uriel and best wishes for many more! 

\appendix
\section{Non-existence of Feynman Measure}

This counter-example appeared in the introduction section in \cite{Cameron1960}.
The following exposition is based on \cite{Klauder2003}. See also
Problem 64 of Chapter X in \cite{RS1975}, \cite{Albeverio1997},
or Chapter 6 and Appendix K in \cite{Strocchi2008}.

All we need to know is the convolution of two Gaussian distributions
$$p_1*p_2(x)=\frac{1}{\sqrt{2\pi(\sigma_1^2+\sigma_2^2)}}\exp\left(-\frac{x^2}{2(\sigma_1^2+\sigma_2^2)} \right)$$
for
$$p_1(x)=\frac{1}{\sqrt{2\pi\sigma_1^2}}\exp\left(-\frac{x^2}{2\sigma_1^2} \right),
p_2(x)=\frac{1}{\sqrt{2\pi\sigma_2^2}}\exp\left(-\frac{x^2}{2\sigma_2^2} \right),$$
where $\sigma_1,\sigma_2 >0.$
For positive $\lambda$ we compute the $N$-fold convolution products for $(N+1)$ functions
by repeated use of it,
$$I=\left( \frac{\lambda}{2\pi \epsilon}\right)^{\frac{N+1}{2}}
\int\ldots\int \exp \left( -\frac{\lambda}{2\epsilon} \sum_{l=0}^N (x_{l+1}-x_l)^2 \right)\Pi_{l=1}^N dx_l
=\left( \frac{\lambda}{2\pi N \epsilon}\right)^{\frac{1}{2}}
\exp \left( -\frac{\lambda}{2 N\epsilon} (x_{N+1}-x_0)^2 \right).$$
This also holds for complex-valued $\lambda$ provided that  $\Re(\lambda) >0$.
If the above formula remains valid in the limit of $N \to \infty$ while $N\epsilon$=const, the path integral
does make sense. Because absolute convergence implies the existence of the integral,
we are led to estimate the majorant
$$|I| \le \left( \frac{|\lambda|}{2\pi \epsilon}\right)^{\frac{N+1}{2}}
\underbrace{\int\ldots\int \exp \left( -\frac{\Re(\lambda)}{2\epsilon} \sum_{l=0}^N (x_{l+1}-x_l)^2 \right)\Pi_{l=1}^N dx_l}_{=J,\; \mbox{say.}}.$$
We have
$$J=\left( \frac{2\pi \epsilon}{\Re(\lambda)}\right)^{\frac{N+1}{2}}\left( \frac{\Re(\lambda)}{2\pi N \epsilon}\right)^{\frac{1}{2}}
\exp \left( -\frac{\Re(\lambda)}{2 N\epsilon} (x_{N+1}-x_0)^2 \right)$$
by an identity obtained by $\lambda \to \Re(\lambda)$ in $I$ above, and hence
$$\left( \frac{\Re(\lambda)}{2\pi \epsilon}\right)^{\frac{N+1}{2}}
\int\ldots\int \exp \left( -\frac{\Re(\lambda)}{2\epsilon} \sum_{l=0}^N (x_{l+1}-x_l)^2 \right)\Pi_{l=1}^N dx_l
=\left( \frac{\Re(\lambda)}{2\pi N \epsilon}\right)^{\frac{1}{2}}
\exp \left( -\frac{\Re(\lambda)}{2 N\epsilon} (x_{N+1}-x_0)^2 \right).$$
Calibrating the prefactors, we find
$$|I| \le
\left( \frac{|\lambda|}{2\pi\epsilon}\right)^{\frac{N+1}{2}}
\left( \frac{2\pi\epsilon}{\Re(\lambda)}\right)^{\frac{N+1}{2}}
\left( \frac{\Re(\lambda)}{2\pi N \epsilon}\right)^{\frac{1}{2}}
\exp \left( -\frac{\Re(\lambda)}{2 N\epsilon} (x_{N+1}-x_0)^2 \right)$$
$$= \left( \frac{|\lambda|}{\Re(\lambda)}\right)^{\frac{N}{2}} \left( \frac{|\lambda|}{2\pi N \epsilon}\right)^{\frac{1}{2}}
\exp \left( -\frac{\Re(\lambda)}{2 N\epsilon} (x_{N+1}-x_0)^2 \right).$$
When $\Im(\lambda)\ne 0,$ we have $\frac{|\lambda|}{\Re(\lambda)}>1.$ Therefore the final line diverges in the limit of
$N \to \infty$ with $N\epsilon$ held fixed.
\enlargethispage{20pt}










\begin{thebibliography}{9}
\bibitem{Hopf1952} Hopf E. 1952 Statistical hydromechanics and functional calculus.
\textit{J. Rat. Mech. Anal.} \textbf{1} 87-123.

\bibitem{HT1953}  Hopf E. and Titt EW. 1953
On certain special solutions of the $\Phi$-equation of statistical hydrodynamics.
\textit{J. Rat. Mech. Anal.} \textbf{2}  587-591.

\bibitem{Neumann1929} von Neumann J. 1929 Beweis des Ergodensatzes und des H-Theorems in der neuen
  Mechanik. \textit{Zeit. Phys.} \textbf{57} 30-70.

\bibitem{Birkhoff1931} Birkhoff GD. 1931. Proof of the ergodic theorem.
\textit{Proc. Nat. Acad. Sci.} \textbf{17} 656-660.

\bibitem{Hopf1937} Hopf E. 1937. \textit{Ergodentheorie}. Berlin: Springer.

\bibitem{TKS2013} Toda M., Kubo R. \& Saito N. 2013.
\textit{Statistical Physics I: Equilibrium Statistical Mechanics}. Berlin: Springer.

\bibitem{Koopman1931} Koopman BO.1931 Hamiltonian systems and transformation in Hilbert space.
\textit{Proc. Nat. Acad. Sci.} \textbf{17} 315.

\bibitem{SM2012}  Singh RK. and Manhas JS. 2012.
\textit{Composition Operators on Function Spaces}. Amsterdam: North Holland.

\bibitem{Weil1932} Weil A. 1932. On systems of curves on a ring-shaped surface.
  \textit{J. Ind. Math. Soc.} \textbf{19} 109--114. Also in \textit{Andr{\'e} Weil Oeuvres Scientifiques Collected Papers
    Volume 1}: 57--62 \& Commentaire (in French) p.522. 1979, Berlin: Springer-Verlag.
\textcolor{black}{
\bibitem{Redei2005} R{\'e}dei M. (ed.) 2005
  \textit{John Von Neumann: Selected Letters}
  Rhode Island: American Mathematical Society.}

\bibitem{Applebaum2019} Applebaum D. 2019.
 \textit{Semigroups of Linear Operators: With Applications to Analysis, Probability and Physics}.
 Cambridge:  Cambridge University Press.

\bibitem{Foias1972} Foias C.  1972 Statistical study of Navier-Stokes equations I.
\textit{Rend. Sem. Mat. Univ. Padova} \textbf{8} 219--348.

\bibitem{Foias1973} Foias C. 1973 Statistical study of Navier-Stokes equations II.
\textit{Rend. Sem. Mat. Univ. Padova} \textbf{49} 9--123.

\bibitem{VF1980} Vishik MI. and Fursikov AV. 1988
\textit{Mathematical problems of statistical hydromechanics}.
Dordrecht and Boston: Kluwer.

\bibitem{Mezic2013} Mezi{\'c} I. 2013. Analysis of fluid flows via spectral properties of the Koopman operator.
\textit{Ann. Rev. Fluid Mech.} \textbf{45} 357--378.

\bibitem{RD2017} Rowley CW. \& Dawson ST. 2017. Model reduction for flow analysis and control.
 \textit{Ann. Rev. Fluid Mech.} \textbf{49} 387--417.

\bibitem{PK2018} 
  Page J. and  Kerswell RR. 2018. Koopman analysis of Burgers equation. \textit{Phys. Rev. Fluids}
  \textbf{3} 071901.

\bibitem{Foias1974}  Foias C. 1974
 A functional approach to turbulence. \textit{Russ. Math. Surv.} \textbf{29} 293.

\bibitem{Edwards1964a}  Edwards SF. 1964
  The statistical dynamics of homogeneous turbulence.
  \textit{J. Fluid Mech.} \textbf{18} 239-273.

\bibitem{Edwards1964b} Edwards SF. 1964
  A new method of solution for quantum field theory and associated problems.
    in \textit{Proceedings of a conference on the theory and applications of analysis
    in function space held at Endicott House in Dedham, Massachusetts, June 9-13, 1963},
  eds. W.T. Martin and I.E. Segal. MIT Press: Cambridge. 31-50

\bibitem{Rosen1960}  Rosen G. 1960
Turbulence Theory and Functional Integration. I. and II.  
\textit{Phys. Fluids} \textbf{3} 519-524 and 525-528.

\bibitem{Rosen1969} Rosen G. 1969
\textit{Formulation of Classical and Quantum Dynamical Theory}.
New York and London: Academic Press.

\bibitem{Novikov1961} Novikov EA. 1961
  The solution of certain equations with variational derivatives. (in Russian)
\textit{Uspekhi Mat. Nauk} \textbf{16} 135--141.

\bibitem{IW2014} Ikeda N. and Watanabe S. 2014
  \textit{Stochastic differential equations and diffusion processes.} Amsterdam: Elsevier.

\bibitem{MY1975} Monin AS. and  Yaglom AM. 1975
\textit{Statistical fluid dynamics: mechanics of turbulence vol.2}
MIT Press.

\bibitem{Tatarskii1962} Tatarskii VI. 1962
Application of the methods of quantum field theory to the problem of 
degeneration of homogeneous turbulence.
\textit{Sov. Phys. JETP} \textbf{15} 961--967.

\bibitem{Klyatskin2015} Klyatskin VI. 2015
  \textit{Stochastic Equations: Theory and Applications in Acoustics, Hydrodynamics,
  Magnetohydrodynamics, and Radiophysics, Volume 1 \& 2}.
Berlin: Springer.

\bibitem{Novikov1965} Novikov EA. 1965 Functionals and the random force method in
turbulence theory. \textit{Sov. Phys. JETP} \textbf{20} 1290--4.

\bibitem{Furutsu1963} Furutsu K. 1963
  On the statistical theory of electromagnetic waves in a fluctuating medium: I.
  \textit{J. Res. Natl. Bur. Stand.} \textbf{67D} 303--23.

\bibitem{Donsker1964} Donsker MD. 1964
  On function space integrals.
    in \textit{Proceedings of a conference on the theory and applications of analysis
    in function space held at Endicott House in Dedham, Massachusetts, June 9-13, 1963},
  eds. W.T. Martin and I.E. Segal. MIT Press: Cambridge. 17-30.

\bibitem{DL1962} Donsker MD. and Lions JL. 1996
Frechet-Volterra variational equations, boundary value problems, and 
function space integrals. \textit{Acta Math.} \textbf{108} 147-228.

\bibitem{Hosokawa1968} Hosokawa I. 1968 A functional treatise on statistical hydromechanics with random force
  action. \textit{J. Phys. Soc. Jpn.} \textbf{25} 271-278.

\bibitem{Frisch1968}  Frisch U. 1968  Wave propagation in random media
in \textit{Probabilistic Methods in Applied Mathematics: volume 1.}
 A.T. Bharucha-Reid (ed.) New York and London: Academic Press.

\bibitem{Zinn-Justin2021}  Zinn-Justin J. 2021.
  \textit{Quantum Field Theory and Critical Phenomena},
  5th edition. Oxford: Oxford University Press.

\bibitem{MSR1973} Martin PC., Siggia ED. and Rose HA., 1973.
Statistical dynamics of classical systems. \textit{Phys. Rev. A} \textbf{8} 423--437.

\bibitem{Wyld1961}  Wyld Jr HW. 1961. Formulation of the theory of turbulence in an incompressible fluid.
\textit{Ann. Phys.} \textbf{14} 143-165.

\bibitem{BdS1998} Botelho LC. and  da Silva EP 1998 The Caldirola-Kanai Theory from "Brownian" Path Integral Quantum
  Mechanics.  \textit{Mod. Phys. Lett. B} \textbf{12} 569--573.

\color{black}
\bibitem{LK1962} Lewis RM. and Kraichnan RH. 1962
A space-time functional formalism for turbulence.
\textit{Commun. Pure Appl. Math.} \textbf{15} 397--411.

\bibitem{Dominicis1976}  de Dominicis C. 1976
Techniques de renormalisation de la th{\'e}orie des champs et dynamique des ph{\'e}nomen{\`e}s critiques.
(in French) \textit{J. Phys. (Paris), Colloq} \textbf{1} 247--253.

\bibitem{Janssen1976}  Janssen H-K. 1976
On a Lagrangean for classical field dynamics and renormalization group
calculations of dynamical critical properties.
\textit{Z. f{\"u}r Physik B} \textbf{23} 377--380.

\bibitem{DM1979} de Dominicis C. and Martin PC. 1979
  Energy spectra of certain randomly-stirred fluids.
  \textit{Phys. Rev. A} \textbf{19} 419--422.

\bibitem{Frisch1995} Frisch U. 1995
  \textit{Turbulence: the legacy of AN Kolmogorov}
  Cambridge: Cambridge University Press.
\color{black}
\bibitem{GM1996} Gurarie V. and Migdal A. 1996 Instantons in the Burgers equation. \textit{Phys. Rev. E} \textbf{54} 4908.

\bibitem{FKLM1996} Falkovich G., Kolokolov I., Lebedev V. and Migdal A. 1996 Instantons and intermittency.
  \textit{Phys. Rev. E} \textbf{54} 4896.


\bibitem{GGS2015}
Grafke T., Grauer R. and  Sch{\"a}fer T. 2015.
The instanton method and its numerical implementation in fluid mechanics.
\textit{J. Phys. A: Math. and Theor.} \textbf{48} 333001.

\bibitem{GGSV2015}
Grafke T., Grauer R., Sch{\"a}fer T. and Vanden-Eijnden, E. 2015.
Relevance of instantons in Burgers turbulence.
\textit{Europhys. Lett.} \textbf{109} 34003.

\bibitem{Rosen1971} Rosen G. 1971.
  Functional calculus theory for incompressible fluid turbulence.
  \textit{J. Math. Phys.} \textbf{12} 812--820.

\bibitem{TR1974} Testa FJ. and Gerald R. 1974
  Theory for incompressible fluid turbulence.
  \textit{J. Franklin Inst.} \textbf{297} 127--133.  

\bibitem{BP1993} Breuer HP. and  Petruccione F. 1993
Burgers's turbulence model as a stochastic dynamical system: Master equation and simulation.
\textit{Phys. Rev. E} \textbf{47} 83.

\bibitem{BP1996a} Breuer HP. and Petruccione F. 1996
  Mesoscopic modelling and stochastic simulations of turbulent flows.
in \textit{Nonlinear Stochastic PDEs: Hydrodynamic Limit and Burgers' Turbulence}
  (ed.) T. Funaki, W.A. Woyczynski, Berlin: Springer. 

\bibitem{BPW1996b}
Breuer HP., Petruccione F. and  Weber F. 1996
On a Fourier space master equation for Navier-Stokes turbulence.
\textit{Z. fur Phys. B} \textbf{100}  461-468.

\bibitem{Kollmann1976}
  Kollmann W. 1976 Functional integrals in the statistical theory of turbulence and the burgers model equation.
  \textit{J. Stat. Phys.} \textbf{14} 291-303.

\bibitem{Hirota1979} Hirota R. 1979 Nonlinear Partial Difference Equations. V. Nonlinear Equations Reducible to Linear Equations. \textit{J. Phys. Soc. Jpn.} \textbf{46} 312--319.

\bibitem{Kollmann2019} Kollmann W. 2019
\textit{Navier-Stokes Turbulence: Theory and Analysis}.
Berlin: Springer.

\bibitem{Venturi2018}  Venturi D. 2018 The numerical approximation of nonlinear functionals and functional
  differential equations. \textit{Phys. Rep.} \textbf{732} 1-102.

  
\bibitem{Botelho2008}  Botelho LCL. 2008
  \textit{Lecture Notes In Applied Differential Equations Of Mathematical Physics}.
  Singapore: World Scientific.

\bibitem{Ohkitani2020} Ohkitani K. 2020
  Study of the Hopf functional equation for turbulence: Duhamel principle and dynamical scaling.
  \textit{Phys. Rev. E} \textbf{101} 013104-1--15.

\bibitem{Cameron1960} Cameron RH. 1960
  A family of integrals serving to connect the Wiener and Feynman integrals.
  \textit{J. Math. Phys.} \textbf{39} 126--140.
  
\bibitem{Klauder2003} Klauder KR. 2003 The Feynman path integral: An historical slice.
in \textit{A Garden of Quanta: Essays in Honor of Hiroshi Ezawa} 55-76.
 

\bibitem{RS1975} Reed M. and Simon B. 1975
  \textit{Methods of Modern Mathematical Physics II: Fourier Analysis, Self-Adjointness}
New York and London:  Academic Press.
  
\bibitem{Albeverio1997} Albeverio S. 1997
  Wiener and Feynman-path integrals and their applications.
  In \textit{Proceedings of Symposia in Applied Mathematics}, Vol. 52, pp. 163-194.
   Rhode Island: American Mathematical Society.

\bibitem{Strocchi2008} Strocchi F. 2008
\textit{An Introduction to the Mathematical Structure of Quantum Mechanics: A Short Course for Mathematicians}. Singapore: World Scientific. 
   
\bibitem{SE2005} Sreenivasan KR. and  Eyink GL. 2005 Sam Edwards and the turbulence theory.
in \textit{Stealing the gold: A celebration of the pioneering physics of Sam Edwards} 66-85.

\bibitem{EF2011} Eyink G. and  Frisch U. 2011 Robert H. Kraichnan.
in \textit{A voyage through turbulence.} (eds.) Davidson P., Kaneda Y., Moffatt HK. and Sreenivasan KR.
Cambridge University Press: Cambridge 329-372.


\end{thebibliography}
\end{document}